\documentclass[aps,prl,twocolumn,superscriptaddress,reprint,showpacs]{revtex4-2}
\usepackage{graphicx}
\usepackage{setspace}
\usepackage{amsmath}
\usepackage{amssymb}
\usepackage{color}
\usepackage{array}
\usepackage{subfigure}
\usepackage{hyperref}
\usepackage{float}
\usepackage{lipsum}

\usepackage[all]{xy}
\newcommand{\RN}[1]{%
  \textup{\uppercase\expandafter{\romannumeral#1}}%
}

\begin{document}
\title{Quantum sensing of low-frequency electric signal enabled by modulated auxiliary field in Rydberg atoms}

\author{Xiayang Fan}
\thanks{Xiayang Fan and Shenchao Jin contributed equally to this work.}
\affiliation{Shanghai Institute of Optics and Fine Mechanics, Chinese Academy of Sciences, Shanghai 201800, China}
\affiliation{University of Chinese Academy of Sciences, Beijing 100049, China}

\author{Shenchao Jin}
\thanks{Xiayang Fan and Shenchao Jin contributed equally to this work.}
\affiliation{Shanghai Institute of Optics and Fine Mechanics, Chinese Academy of Sciences, Shanghai 201800, China}
\affiliation{University of Chinese Academy of Sciences, Beijing 100049, China}

\author{Jiatian Liu}
\affiliation{Shanghai Institute of Optics and Fine Mechanics, Chinese Academy of Sciences, Shanghai 201800, China}
\affiliation{University of Chinese Academy of Sciences, Beijing 100049, China}

\author{Jialiang Zhang}
\affiliation{Qianyuan Laboratory, Hangzhou 310024, China}

\author{Qichao Qi}
\affiliation{Qianyuan Laboratory, Hangzhou 310024, China}

\author{Yuan Sun}
\email[email: ]{yuansun@siom.ac.cn}
\affiliation{Shanghai Institute of Optics and Fine Mechanics, Chinese Academy of Sciences, Shanghai 201800, China}
\affiliation{University of Chinese Academy of Sciences, Beijing 100049, China}
\affiliation{Qianyuan Laboratory, Hangzhou 310024, China}

\begin{abstract}
Rydberg atoms have emerged as a versatile and efficient platform for high-sensitivity quantum sensing of free-space electric fields, with remarkable progress in detecting low-frequency signals. 
To date, low-frequency Rydberg receivers have relied on a constant bias field, typically realized via intra-cell electrodes or Rydberg plasmas generated by photoelectric effects or inter-atomic interactions. 
While these approaches improve sensitivity, they suffer from inherent challenges in calibration, long-term stability, and robustness, hindering practical deployment. 
Here, we propose, design, and experimentally demonstrate a quantum sensing scheme for low-frequency electric signals using modulated auxiliary fields in Rydberg atoms. 
Unlike conventional methods that employ external DC electric fields that are often fully shielded by adsorbed atom layers on the cell walls, we introduce an AC-field modulation strategy. 
The incoming low-frequency signal mixes with the auxiliary field, and together they induce Stark shifts of the Rydberg level. 
These shifts are mapped onto the probe laser via electromagnetically induced transparency (EIT), in a manner analogous to heterodyne detection. 
We demonstrate a sensitivity of $7.5 \pm 2.6~\mathrm{\mu V/(cm\cdot Hz^{1/2})}$ at 5~kHz and a minimal detectable field of $0.26 \pm 0.04~\mathrm{\mu V/cm}$ with an integration time of 1000~s.
Furthermore, we extend this approach to systematically analyze the performance of generalized auxiliary fields containing multiple frequency components. 
By virtue of modulated auxiliary field and quantum frequency mixing, our results establish a robust and systematic framework for quantum sensing of low-frequency electric fields with Rydberg atoms, offering improved sensitivity, stability, and immunity to environmental drifts. 
\end{abstract}
\maketitle

Low-frequency (LF) electric-field sensing has significant applications in space science \cite{berthelier_ice_2006}, geophysical surveys \cite{wang_assessment_2022}, and communication in complex environments \cite{latypov_compact_2022}.
Rydberg atoms have proven to be an exceptionally broadband and sensitive probe of free-space electric fields \cite{Holloway2024Review, Zhang2024Review}, with demonstrated quantum sensing operation spanning DC \cite{osterwalder1999PRL, PhysRevA.84.033402, facon_sensitive_2016}, LF \cite{JauCarter2020PRapp, holloway_electromagnetically_2022, jin2025cpl, FuYunqi2026npj}, radio frequency (RF) \cite{fan_atom_2015, kumar_rydberg-atom_2017, simons_embedding_2019, Zhanglinjie2024OL, Cox2024APL, Cox2024PRapp}, microwave (MW) \cite{Shaffer2012NP, Cox2018PRL, ZhangLinjie2020NP, Yanhui2024SA, LiLin2026PRL, Cox2023PRapp, dds2026arxiv}, and Terahertz (THz) \cite{wade_real-time_2017, downes_full-field_2020, chen_terahertz_2022, krokosz2025Optica, wang_improving_2026} frequencies.
In the LF regime, Rydberg sensing of electric fields $E$ typically relies on the second-order Stark effect, with the energy shift $\delta \propto E^2$.
To achieve high sensitivity, we can introduce an additional DC electric field $E_\mathrm{aux}$ and superimpose it with the LF signal field $E_\mathrm{sig}$, yielding $\delta \propto (E_\mathrm{aux} + E_\mathrm{sig})^2 = 2E_\mathrm{aux}E_\mathrm{sig} + \cdots$.
Conventional implementations of this DC auxiliary field rely on intra-cell electrodes \cite{grimmel_measurement_2015, ma_dc_2020, holloway_electromagnetically_2022}, Rydberg plasmas produced by photoelectric effects or inter-atomic interactions \cite{JauCarter2020PRapp, jin2025cpl, FuYunqi2026npj}.
While effective in boosting the signal, these approaches suffer from persistent issues of calibration drift, limited long-term stability, and environmental sensitivity, which restrict their practical deployment outside the laboratory.
These observations motivate us to consider the potential choices of biasing strategies beyond static fields.

In quantum sensing of RF, MW and THz signals via Rydberg atoms, oscillating auxiliary fields are widely adopted and two basic categories of methods have proved effective.  
The first category utilizes a strong off-resonant RF auxiliary field, which dresses the Rydberg states with Floquet sidebands \cite{miller_radio-frequency-modulated_2016, clark_interacting_2019} and enables a drastic extension of the instantaneous detection bandwidth \cite{cui_extending_2023, rotunno_detection_2023, Zhaojianming2024APL}.
The second category exploits a local-oscillator field detuned close to the signal frequency, shifting the signal information to an intermediate frequency and often delivering a substantial sensitivity gain with a heterodyne scheme \cite{simons_embedding_2019, ZhangLinjie2020NP, jia_span_2021, manchaiah_probing_2026, liang_exceptional_2026}.
Both approaches circumvent the need for internal electrodes and have become the standard  routines in MW/RF Rydberg electrometry. 
On the other hand, in nitrogen-vacancy (NV) center based RF field sensing, a quantum frequency mixing (QFM) effect was recently proposed to extend the sensing ability through the nonlinear response of the NV ensemble \cite{wang_sensing_2022, yin_high-resolution_2025, wangya2025nature}.
Nevertheless, these previously established methods cannot be directly applied to LF signals because the underlying physical mechanism is distinctly different.
Facing this challenge, we raise the question whether an AC auxiliary field can realize improvements like QFM in Rydberg atoms for LF sensing, combining an electrode-free design with quadratic Stark cross-term amplification.

\begin{figure*}[htbp]
\centering
\includegraphics[width=0.8\textwidth]{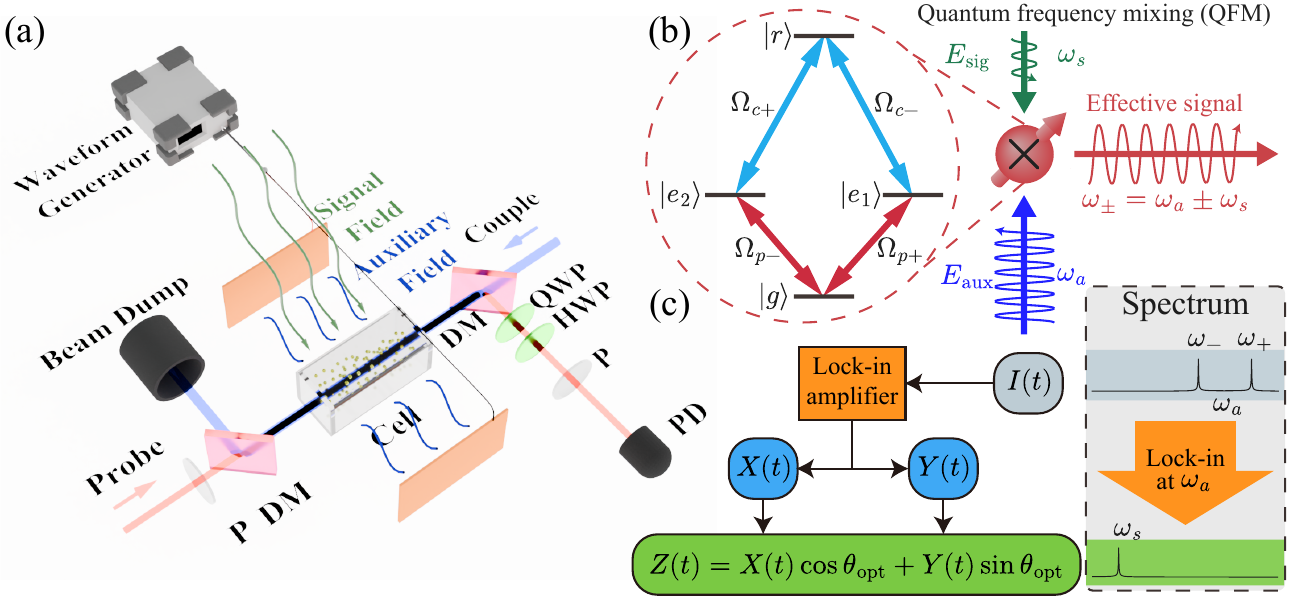}
\caption{(a) Schematic diagram of the proposed approach of modulated auxiliary field with weak-measurement-enhanced readout. In our experiment, the auxiliary field and the signal field are both applied to the copper plates to simulate real-world application scenarios \cite{SuppInfo}. (b) Principles of the modulated auxiliary field and QFM, including the relevant atomic levels and transitions. (c) Signal reconstruction process. P: polarizer, DM: dichroic mirror, HWP: half-wave plate, QWP: quarter-wave plate, PD: photodiode.}\label{fig1}
\end{figure*}

To address the aforementioned critical question, we design and experimentally realize a QFM-type method for quantum sensing of LF electric-field with Rydberg atoms, combining the modulated auxiliary field with a weak-measurement-enhanced readout, all implemented in a conventional uncoated glass vapor cell.
More specifically, a locally generated AC auxiliary field is superimposed with the incoming LF signal field,
and together, they induce a second-order Stark shift on the Rydberg states, producing a cross term proportional to the product of the signal and auxiliary amplitudes. 
This mixing amplifies the weak LF signal and shifts it to sidebands of the auxiliary field, enabling signal extraction at a higher frequency which helps to improve noise-resilience.
The resulting enhanced energy shift on the Rydberg state is encoded onto the change of probe laser polarization \cite{PhysRevA.97.023815} through the dispersive response of the Rydberg electromagnetically induced transparency (EIT), while the weak-measurement protocol suppresses technical fluctuations during readout.
Weak measurement has long been recognized as a powerful tool in quantum sensing \cite{aharonov_how_1988, dressel_colloquium_2014, song_weak_2023, wang__enhanced_2025}, and has recently been introduced into sensing LF signal via Rydberg atoms \cite{PhysRevA.97.023815, wang2026arxiv}.
The proposed method was successfully implemented experimentally, yielding a sensitivity of $7.5 \pm 2.6~\mathrm{\mu V/(cm\cdot Hz^{1/2})}$ at 5~kHz and a minimal detectable field of $0.26 \pm 0.04~\mathrm{\mu V/cm}$ for an integration time of 1000~s.
This technique requires no internal electrodes and is intrinsically stable, calibration-friendly, and robust against environmental drifts, offering a practical route toward deployable LF electric-field sensors.

The operational principle and experimental schematic of the proposed method are depicted in Fig.~\ref{fig1}.
Fig.~\ref{fig1}(a) schematically depicts the experimental setup, which is built around an uncoated, isotopically enriched $^{87}$Rb glass vapor cell of dimensions $25~\mathrm{mm}\times 25~\mathrm{mm}\times 50~\mathrm{mm}$, operated at room temperature.
Two frequency-stabilized lasers traverse the cell in a counter-propagating geometry:
a 780~nm probe laser, focused to a waist of 0.3~mm with a Rabi frequency $\Omega_p = 2\pi \times 18~\mathrm{MHz}$, and a 480~nm coupling laser with a waist of 0.4~mm and $\Omega_c = 2\pi \times 2.2~\mathrm{MHz}$. 
Both lasers are locked to an ultra-low expansion (ULE) reference cavity \cite{jin2025cpl, SuppInfo}.
A pair of parallel copper plates ($300~\mathrm{mm}\times 300~\mathrm{mm}$, spacing 56~mm), driven by an arbitrary waveform generator that produces the combined signal and auxiliary fields, generates a uniform electric field at the position of the atoms.
The entire assembly---vapor cell and field plates---is housed within a three-axis Helmholtz coil system that actively cancels the ambient magnetic field and applies a static bias field of 5.0~Gs along the laser propagation direction to lift the energy degeneracy.
We utilize the polarization degree of freedom and the weak measurement readout protocol introduced in Ref.~\cite{wang2026arxiv}, with the simplified level diagram shown in Fig.~\ref{fig1}(b).
The probe laser is initially prepared in the diagonal polarization state $|\phi_i\rangle = (|H\rangle + |V\rangle)/\sqrt{2}$, corresponding to linear polarization at $\pi/4$ relative to the $x$-axis.
Together with the coupling laser, it drives $^{87}$Rb atoms in a room-temperature vapor cell.
In Fig.~\ref{fig1}(b), the probe laser couples the ground state $|g\rangle \in |5^{2}S_{1/2}, F=2\rangle$ to the two Zeeman sublevels $|e_{1,2}\rangle$ of the excited state $|5^{2}P_{3/2}, F=3\rangle$, while the coupling laser excites the atoms from $|e_{1,2}\rangle$ to the Rydberg state $|r\rangle \in |70^{2}D_{5/2}, m_j=5/2\rangle$.

Simultaneously, an LF signal field $E_\mathrm{sig}$ and an auxiliary AC field $E_\mathrm{aux}$ at a different but higher frequency are applied to the interaction region.
These two fields jointly affect the Rydberg state $|r\rangle$ via the Stark effect, producing a time-dependent energy shift $\delta$ that embodies QFM.
Generally speaking, $E_\mathrm{sig}$ is relatively weak, and the locally generated $E_\mathrm{aux}$ predominantly determines the polarization of the total electric field that drives the atomic response.
Thus, the method inherently operates with a pre-selected polarization.
Through the dispersive response of the EIT medium, $\delta$ is mapped onto a polarization change of the probe light.
After the probe passes through a post-selection polarizer set to an angle $-\pi/4 + \varepsilon$, the transmitted intensity is given by \cite{wang2026arxiv, SuppInfo}
\begin{equation}\label{Eq_wmI}
  I(t) \approx I_0 \sin^2 \varepsilon \left( 1 + \beta \mathrm{Im} A_w \cdot \delta \right),
\end{equation}
where $I_0$ is the input intensity, $A_w = i \cot \varepsilon$ is the weak value, and $\beta$ is a coefficient that encapsulates the atomic response 
$\beta = \mathcal{A} / \Gamma_p \cdot 2(\Omega_{c-}^2 - \Omega_{c+}^2) e^{-\mathcal A} / [2\Omega_{c}^2 - (\Omega_{c-} + \Omega_{c+})^2 (1 - e^{-2 \mathcal A})]$.
In this expression, $\mathcal{A}$ is the total absorption on the probe transition, $\Gamma_p$ is the power-broadened linewidth of the EIT feature, $\Omega_{c\pm}$ are the Rabi frequencies for the $\sigma^+$ and $\sigma^-$ polarization components of the coupling laser, and $\Omega_c^2 = \Omega_{c-}^2 + \Omega_{c+}^2$.
As established in Ref.~\cite{wang2026arxiv}, this weak-measurement protocol not only translates the Rydberg energy shift into an intensity signal but also suppresses technical noise via the post-selection angle and enhances robustness against fluctuations of the two-photon resonance.

For the processes under consideration, the energy shift of the Rydberg state is dominated by the second-order Stark effect \cite{zimmerman_stark_1979} as $\delta = -\alpha E_\mathrm{tot}^2 /2$, where $\alpha = 2439.5~\mathrm{MHz \cdot cm^2 \cdot V^{-2}}$ is the polarizability of $|r\rangle$ numerically calculated with the Alkali Rydberg Calculator (ARC) \cite{arc}.
The total external electric field $E_\mathrm{tot}$, aligned along the polarization of $E_\mathrm{aux}$, is the linear superposition of $E_\mathrm{aux}$ and $E_\text{sig}$, 
\begin{equation}\label{Eq_Etot}
E_\mathrm{tot} = E_a \cos(\omega_a t + \phi_a) + E_s \cos(\omega_s t + \phi_s), 
\end{equation}
where $E_{a, s}$, $\omega_{a, s}$, and $\phi_{a, s}$ denote the amplitudes, frequencies, and phases of the auxiliary and signal fields, respectively.
Expanding the square in $\delta$ leads to the QFM spectrum
\begin{equation}\label{Eq_ftot}
\delta =  \delta_\mathrm{dc} + \delta_\mathrm{cr} + \delta_\mathrm{ot},
\end{equation}
with a quasi-static term 
$\delta_\mathrm{dc} = -\alpha (E_a^2 + E_s^2) /4$, 
a second-harmonic term 
$\delta_\mathrm{ot} = -\alpha [ E_a^2 \cos(2\omega_a t + 2\phi_a) + E_s^2 \cos(2\omega_s t + 2\phi_s) ] /4$ 
and, most importantly, the cross term $\delta_\mathrm{cr} = -\alpha E_a E_s [\cos(\omega_+ t + \phi_+) + \cos(\omega_- t + \phi_-)] / 2$,
with the sideband frequencies $\omega_\pm = \omega_a \pm \omega_s$ and phases $\phi_\pm = \phi_a \pm \phi_s$.
The cross term constitutes the QFM signal: it is linearly proportional to the signal amplitude $E_s$ and its magnitude is scaled by the auxiliary amplitude $E_a$. 
Therefore, the use of the AC auxiliary field imprints the complete information of $E_\mathrm{sig}$ onto the sidebands of $\delta$.

\begin{figure}
\centering
\includegraphics[width=0.45\textwidth]{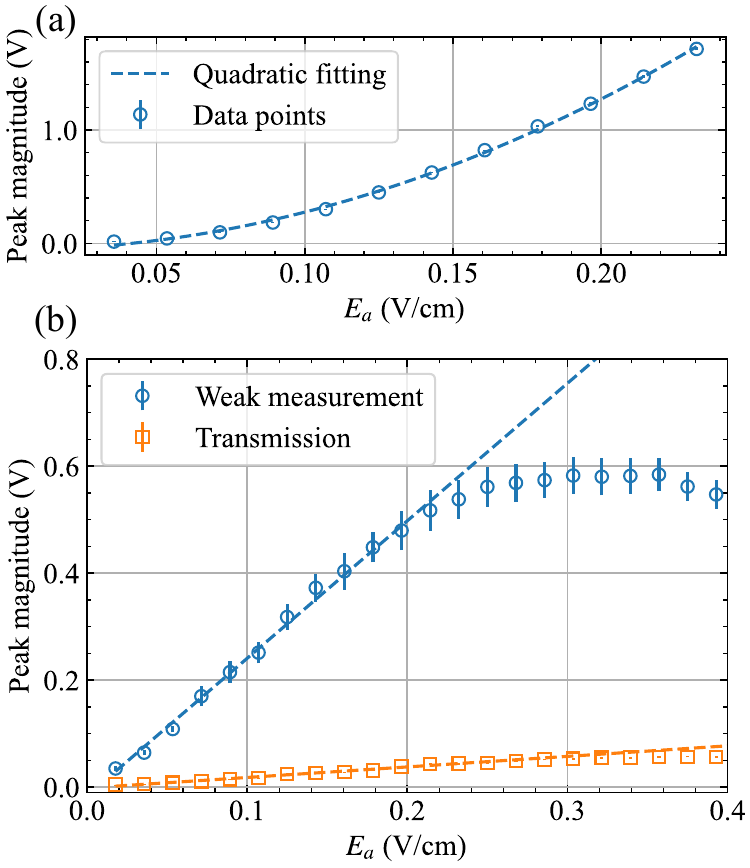}
\caption{Demodulated signal magnitude as a function of the auxiliary field amplitude $E_a$, demodulated at (a) $2\omega_a$ and (b) $\omega_-$. Error bars represent one standard deviation from five independent experiments. Dashed lines show a quadratic fit in (a) and a linear fit in (b).}\label{fig2}
\end{figure}

As an initial verification of the QFM decomposition, we experimentally confirm Eq.~\ref{Eq_ftot} using the data shown in Fig.~\ref{fig2}.
We apply the combined field $E_\mathrm{tot}$ with $\omega_a = 2\pi\times 25~\mathrm{kHz}$, $\omega_s = 2\pi\times 5~\mathrm{kHz}$, and $E_s = 18~\mathrm{mV/cm}$, and feed the photodiode signal into a lock-in amplifier, then the demodulated magnitude $R$ is recorded.
Since the $\delta_\mathrm{dc}$ term shifts the two-photon resonance, we scan the coupling laser frequency at each $E_a$ and take the peak of the resulting magnitude-versus-frequency curve.
When the lock-in is referenced to $2\omega_a$, the peak magnitude exhibits a clear quadratic dependence on $E_a$ [Fig.~\ref{fig2}(a)], confirming the $\delta_\mathrm{ot}$ term.
We then demodulate at the difference frequency $\omega_-$ to isolate the cross term $\delta_\mathrm{cr}$.
As shown in Fig.~\ref{fig2}(b), the magnitude first increases linearly with $E_a$, consistent with $\delta_\mathrm{cr} \propto E_a$.
However, the signal does not increase indefinitely, it decreases beyond an optimal value of $E_a$.
This non-monotonic behavior arises because the linear mapping between the detected intensity $I$ and the energy shift $\delta$ holds only for shifts much smaller than the EIT linewidth.
When the modulation of $\delta_\mathrm{cr}$ becomes comparable to the linewidth, the window of dispersive response saturates.
We therefore operate at $E_a = 0.32~\mathrm{V/cm}$ in all subsequent measurements, experimentally maximizing the signal-to-noise ratio (SNR).

Unlike experiments that rely on the linear Stark effect \cite{jin2025cpl}, our approach reconstructs the signal field $E_\mathrm{sig}$ from the QFM sidebands encoded in the measured light intensity $I(t)$.
As illustrated in Fig.~\ref{fig1}(c), we demodulate $I(t)$ at the auxiliary frequency $\omega_a$ and record the in-phase $X(t)$ and quadrature $Y(t)$ components.
From these quadratures we arrive at an optimal linear combination $Z(t) = X(t) \cos \theta + Y(t) \sin \theta$, where the phase $\theta$ is chosen to maximize the signal at $\omega_s$.
The optimal phase angle is given analytically by
$\cos \theta_\mathrm{opt} = \sqrt{(r+u)/(2r)}$, $\sin \theta_\mathrm{opt} = \mathrm{sgn}(v) \sqrt{(r-u)/(2r)}$, 
with $u = |X(\omega_s)|^2 - |Y(\omega_s)|^2$, $v = 2 \mathrm{Re}(X(\omega_s)Y^*(\omega_s))$, $r = \sqrt{u^2 + v^2}$, and $\mathrm{sgn}(v)$ the sign of $v$ \cite{SuppInfo}.
Eventually, the reconstructed signal takes the simple form 
$Z(t) \propto E_a \cos (\omega_s t + \phi_s + \Delta \phi)$, 
where $\Delta \phi$ is a constant phase offset arising from the circuit loop and post-processing.
In this way, the full information of amplitude and phase of the signal field is faithfully recovered.

This approach differs fundamentally from the AC-assisted methods commonly employed in the Rydberg electrometry at MW and RF  frequencies \cite{simons_embedding_2019, ZhangLinjie2020NP, jia_span_2021, manchaiah_probing_2026, liang_exceptional_2026}.
In those heterodyne protocols, the intensity signal is mixed down to the intermediate frequency $\omega_- = \omega_a - \omega_s$, and a low-pass filter extracts a time-averaged amplitude; 
the information carried by the sum-frequency sideband $\omega_+ = \omega_a + \omega_s$ is discarded, incurring an inherent $\sim3~\mathrm{dB}$ loss in sensitivity.
In contrast, our QFM method reconstructs the original waveform $Z(t)$ directly at the signal frequency $\omega_s$, preserving both amplitude and phase and thereby achieving a complete, coherent detection of the LF electric field.

\begin{figure}[t]
\centering
\includegraphics[width=0.45\textwidth]{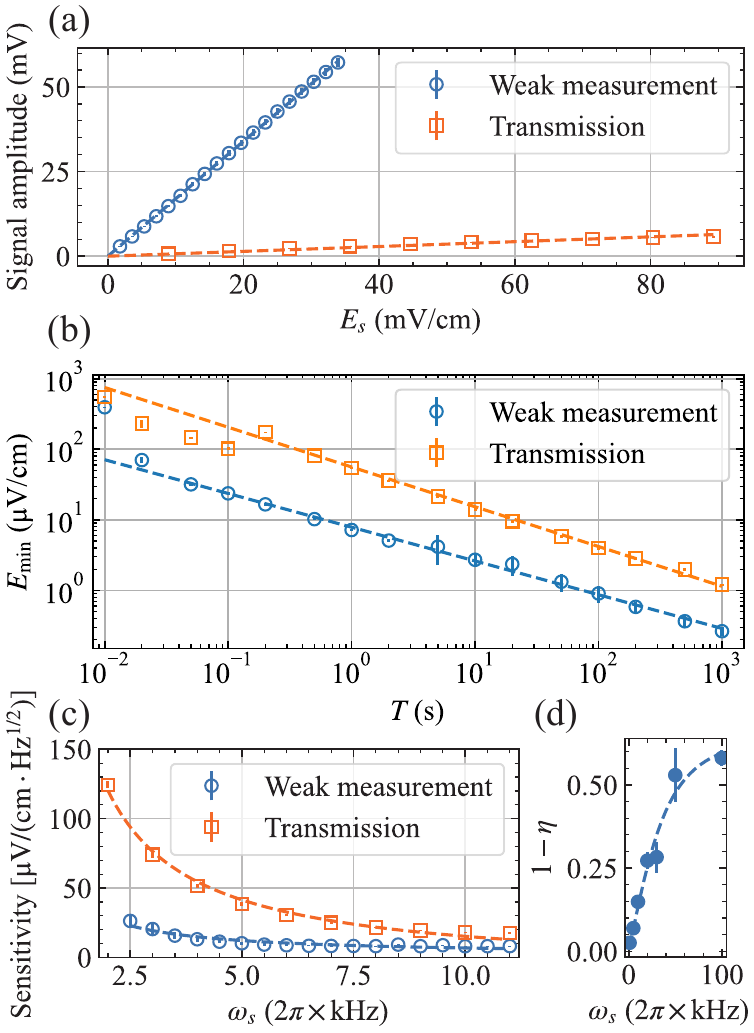}
\caption{(a) Reconstructed signal amplitude as a function of the signal electric field amplitude. 
(b) Minimal detectable field versus integration time. 
(c) Sensitivity as a function of signal frequency. 
(d) Experimentally measured frequency dependent screening ratio $\eta$.
Error bars represent one standard deviation from five independent measurements.}\label{fig3}
\end{figure}

The main experimental results are summarized in Fig.~\ref{fig3}.
We first verify the linear response of the reconstructed signal amplitude $|Z|$ at $\omega_s$ as a function of $E_s$. 
As shown in Fig.~\ref{fig3}(a), the data follow a clear linear relation $|Z| = k E_s$ with $k = 1.698 \pm 0.003~\mathrm{cm}$, confirming faithful transduction of the LF electric field via QFM.
To determine the sensitivity, we obtain the noise floor $N_\mathrm{tot}$ from a frequency spectrum with a resolution bandwidth of $10~\mathrm{Hz}$ over a $\pm 700~\mathrm{Hz}$ window around $\omega_s$, corresponding to an integration time $T = 0.1~\mathrm{s}$.
The sensitivity is then given by $\mathcal{S} = N_\mathrm{tot} / k \cdot \sqrt{T} = 7.5 \pm 2.6~\mathrm{\mu V/(cm\cdot Hz^{1/2})}$.

To assess the weak-field sensing capability and system stability, we perform long-term integration measurements, as shown in Fig.~\ref{fig3}(b).
The experimentally measured minimal detectable field for an integration time $T$ is given by $E_\mathrm{min} = N_\mathrm{tot} / k$.
As $T$ increases, $E_\mathrm{min}$ steadily decreases, reaching $0.26 \pm 0.04~\mathrm{\mu V/cm}$ at $T = 1000~\mathrm{s}$.
A power-law fit to the data yields $E_\mathrm{min} \propto T^{-0.48}$, whose exponent agrees well with the quantum-noise-limited scaling $T^{-0.5}$. 
The slight discrepancy indicates the presence of excess technical noise in our system.

We calibrate the frequency-dependent screening of the vapor cell using a procedure similar to Ref.~\cite{Lim2023APL}, with the results shown in Fig.~\ref{fig3}(d).
At 5~kHz, only $6.9\%$ of the externally applied field reaches the atoms \cite{SuppInfo}, and the remainder is attenuated by the cell walls.
Correcting for this, the intrinsic sensitivity seen by the atoms inside the cell at 5~kHz becomes $\mathcal{S} = 0.52~\mathrm{\mu V/(cm\cdot Hz^{1/2})}$, and the minimal detectable field at 1000~s becomes $E_\mathrm{min} = 18~\mathrm{nV/cm}$.
The sensitivity of measurement as a function of signal frequency $\omega_s$ is presented in Fig.~\ref{fig3}(c).
As expected from the screening effect, lower frequencies are more strongly suppressed, leading to a degraded sensitivity.
Nevertheless, even in our uncoated glass cell, the scheme still delivers a sensitivity of $26~\mathrm{\mu V/(cm\cdot Hz^{1/2})}$ at 2.5~kHz.

To further benchmark the modulated auxiliary field technique, we also test a conventional transmission configuration, in which the polarizer in front of the photodiode is removed.
As shown in Figs.~\ref{fig2}(b) and \ref{fig3}, the transmission readout exhibits the same qualitative behavior as the weak-measurement-assisted polarization detection, but its sensitivity is substantially reduced, highlighting the advantage of the weak measurement scheme.

\begin{figure}[htbp]
\centering
\includegraphics[width=0.4\textwidth]{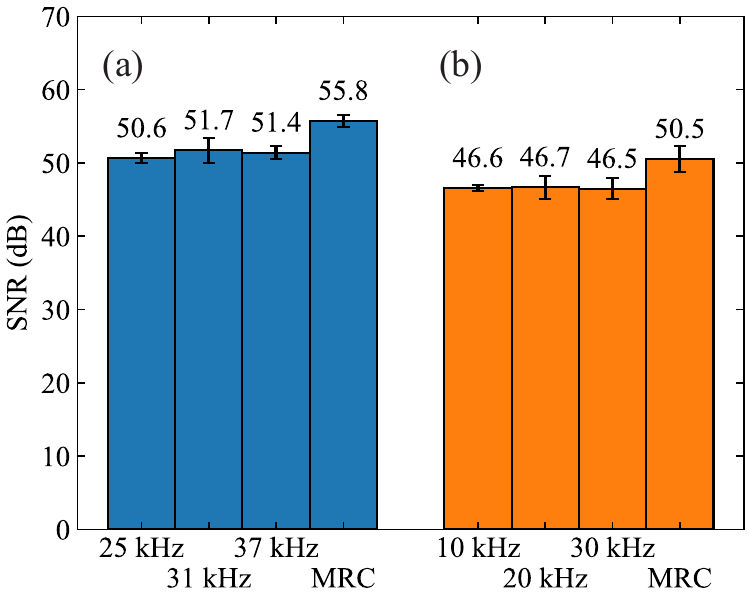}
\caption{SNR histograms for the individual auxiliary frequency components and their MRC combination. 
Data are shown for auxiliary tones at (a) 25~kHz, 31~kHz and 37~kHz (common amplitude of 0.27~V/cm); (b) 10~kHz, 20~kHz and 30~kHz (common amplitude of 0.18~V/cm). 
Error bars represent one standard deviation from five independent measurements.}\label{fig4}
\end{figure}

The modulated auxiliary field is by no means limited to a single tone.
To fully explore the potential of this approach, we extend it to multi-frequency QFM.
Then the cross term generalizes to
\begin{align}\label{Eq_multifreq}
\delta_\mathrm{cr} = -\frac{\alpha E_s}{2} \sum_i E_{a,i} [\cos(\omega_{+,i} t + \phi_{+,i}) \nonumber\\
+ \cos(\omega_{-,i} t + \phi_{-,i})],
\end{align}
with subscript $i$ representing the $i$-th frequency component of the auxiliary fields.
Since the intensity signal now contains contributions at multiple auxiliary frequencies, we replace the hardware lock-in amplifier with a high-speed data acquisition card, which enforces numerical lock-in detection and signal reconstruction for each auxiliary frequency component separately.
After optimizing the amplitudes and phases of these components \cite{SuppInfo}, we obtain the results shown in Fig.~\ref{fig4}.
The histograms compare the SNR of the individual frequency components with that of their combination, obtained via the standard maximal ratio combining (MRC) method \cite{alamouti_simple_1998, SuppInfo}.
We find that the MRC-combined signal consistently yields a higher SNR than any single-frequency component, regardless of whether the cross-term sidebands are well separated (auxiliary fields at 25~kHz, 31~kHz and 37~kHz) or partially overlapping (auxiliary fields at 10~kHz, 20~kHz and 30~kHz). 
This demonstrates that a multi-frequency auxiliary field can, in principle, provide better sensitivity than the single-frequency case.
Thus, the general concept of a modulated auxiliary field can substantially enhance quantum sensing of LF electric signals with Rydberg atoms, enabling a rich variety of modulation strategies.

As discussed in connection with Fig.~\ref{fig2}(b), the finite bandwidth of the Rydberg EIT is the primary factor limiting further sensitivity enhancement.
When the modulation of $\delta_\mathrm{cr}$ induced by $E_\mathrm{aux}$ sweeps over a range exceeding this bandwidth, the transmitted intensity ceases to be a faithful sinusoidal replica of the applied field. 
Instead, it becomes a train of pulses, whose duty cycle shrinks with increasing $E_a$.
The resulting reduction in the Fourier component at the cross frequency $\omega_\pm$ ultimately caps the usable auxiliary field strength.
Aiming at alleviating this restriction, we apply a compensating frequency modulation to the coupling laser \cite{SuppInfo}.
By modulating the coupling frequency synchronously with the auxiliary-field-induced Stark shift, the effective two-photon detuning experienced by the atoms is actively confined within the EIT linewidth even for large auxiliary fields, thereby restoring a linear and efficient transduction of the signal field.
Practical imperfections of the modulation source---in particular its instability and limited modulation range---currently prevent us from fully translating this scheme into higher sensitivity.
Nevertheless, the strong measured dependence of the SNR on the relative phase between the frequency modulation and the auxiliary field \cite{SuppInfo} clearly demonstrates the potential of this approach to overcome the bandwidth bottleneck.


Nevertheless, the present performance remains well above the standard quantum limit (SQL) \cite{Cox2018PRL, Yanhui2024SA}.
In addition to the strong screening of LF fields by the uncoated glass cell, our system is affected by residual Doppler broadening from room-temperature atomic motion and by collisional decoherence, all of which broaden the EIT linewidth and limit the effective signal transduction.
Looking forward, extending the approach of modulated auxiliary field to two-component or multi-species atomic ensembles \cite{Yuan2024SCPMA} appears as a possible solution.
Moreover, the QFM process demonstrated here can serve as a programmable analog processing stage prior to digitization by analog-to-digital converters (ADCs).
In machine-learning-assisted precision sensing, the analog pre-processing is believed to substantially enhance optimization, enabling higher sensing accuracy even at lower ADC rates \cite{del_hougne_learned_2020, li_intelligent_2020, gao_programmable_2023, zhang_system--chip_2024, senanian_microwave_2024, xu_analog_2024, santra_machine_2025, wu_microring_2026}.
Therefore, our results can hopefully help to facilitate the integration of machine learning techniques with Rydberg atomic sensors \cite{liu_deep_2022, ji_local_2025, kang_deepq-mimo_2026}.

In summary, we have introduced and experimentally validated a QFM method for low-frequency electric-field sensing with Rydberg atoms, integrated with a weak-measurement-enhanced polarization readout.
By replacing the conventional static bias with a modulated auxiliary field and exploiting the quadratic Stark nonlinearity, our scheme circumvents the need for intra-cell electrodes and the associated instabilities, while simultaneously amplifying weak LF signals through the mixing cross term.
The combination of QFM, optimal phase-sensitive reconstruction, and weak-measurement-based noise suppression yields a sensitivity of $7.5 \pm 2.6~\mathrm{\mu V/(cm\cdot Hz^{1/2})}$ at 5~kHz and a minimal detectable field of $0.26 \pm 0.04~\mathrm{\mu V/cm}$ after 1000~s of integration.
After accounting for the LF screening effect of the cell, the intrinsic sensitivity seen by the atoms inside the cell at 5~kHz becomes $0.52~\mathrm{\mu V/(cm\cdot Hz^{1/2})}$, and the minimal detectable field at 1000~s becomes $18~\mathrm{nV/cm}$.
We have further demonstrated that the concept naturally extends to multi-frequency QFM with MRC.
We believe that the framework established here provides a versatile and robust foundation for next-generation LF electric-field quantum sensors.

The authors gratefully acknowledge funding supports from the National Key R\&D Program of China (Grant No. 2024YFB4504002), the Strategic Priority Research Program of the Chinese Academy of Sciences (Grant No. XDB1690300), and the Space Application System of China Manned Space Program. Shenchao Jin acknowledges the support from the China Postdoctoral Science Foundation under Grant No. 2024M753359.


\bibliography{xylophone_ref}
\end{document}